# Quantum Algorithmic Gate-Based Computing: Grover Quantum Search Algorithm Design in Quantum Software Engineering

Ulyanov Sergey V.[*] and Ulyanov Viktor S.[†]

[*]Institute of System Analysis and Management, Dubna State University
[*]Meshcheryakov Laboratory of Information Technologies, Joint Institute for Nuclear Research
[†]Department of Information Technologies, Moscow State University of Geodesy and Cartography (MIIGAiK)
[*]Email: srg.v.ulyanov@gmail.com
[†]Email: ulyanovik@gmail.com

**Abstract**

The difference between classical and quantum algorithms (QA) is following: problem solved by QA is coded in the structure of the quantum operators. Input to QA in this case is always the same. Output of QA says which problem coded. In some sense, give a function to QA to analyze and QA returns its property as an answer without quantitative computing. QA studies qualitative properties of the functions. The core of any QA is a set of unitary quantum operators or quantum gates. In practical representation, quantum gate is a unitary matrix with particular structure. The size of this matrix grows exponentially with an increase in the number of inputs, which significantly limits the QA simulation on a classical computer with von Neumann architecture. Quantum search algorithm (QSA) - models apply for the solution of computer science problems as searching in unstructured data base, quantum cryptography, engineering tasks, control system design, robotics, smart controllers, etc. Grover's algorithm is explained in details along with implementations on a local computer simulator. The presented article describes a practical approach to modeling one of the most famous QA on classical computers, the Grover algorithm.

## 1 Introduction: Applied Quantum Search Algorithm Model

Grover Quantum Search Algorithm (QSA) is one of the famous quantum algorithms (QA) that outperform their classical counterparts [1-4]. In the conventional linear search algorithm, it required $\mathcal{O}(N)$ comparisons to find an element in an array of length $N$. Grover's algorithm achieves a quadratic speed up; i.e., it has a complexity of $\mathcal{O}(\sqrt{N})$. Grover's search algorithm provides an example of the speed-up that would be offered by quantum computers (if and when they are built) and has the important application in solution of *global optimization* control problems. The problem solved by Grover's algorithm is finding a sought-after («*marked*») element in an unsorted database (DB) of size $N$. To solve this problem, a classical computer would need $\frac{N}{2}$ database queries on average, and in the worst case it would $N-1$ queries.

Thus, using Grover's algorithm, a quantum computer can find the marked state using only $\mathcal{O}(\sqrt{N})$ quantum data queries. In the case of $M$ «*marked*» elements in an unsorted *DB* of size $N$ speed-up of quantum search process increases as $\mathcal{O}\left(\sqrt{\frac{N}{M}}\right)$. It is believed that this complexity is optimal. This speed up is inherently due to the parallel computational nature of quantum operators that can affect all of the coefficients in the state expansion at once.

## 1.1 General design structure of quantum algorithms

A quantum algorithm (QA) calculates the qualitative properties of the function $f$.

From a mathematical standpoint, a function $f$ is the map of one logical state into another. The problems solved by a QA can be stated as follows:

> Given a function $f:\{0,1\}^n \to \{0,1\}^m$; find a certain property of the function $f$.

Or in the symbolic form as:

| Input | A function $f: \{0,1\}^n \to \{0,1\}^m$ |
|---|---|
| Problem | Find a certain property of $f$ |

Figure 1 is a block diagram showing a gate approach for simulation of a QA using classical computers [5]:

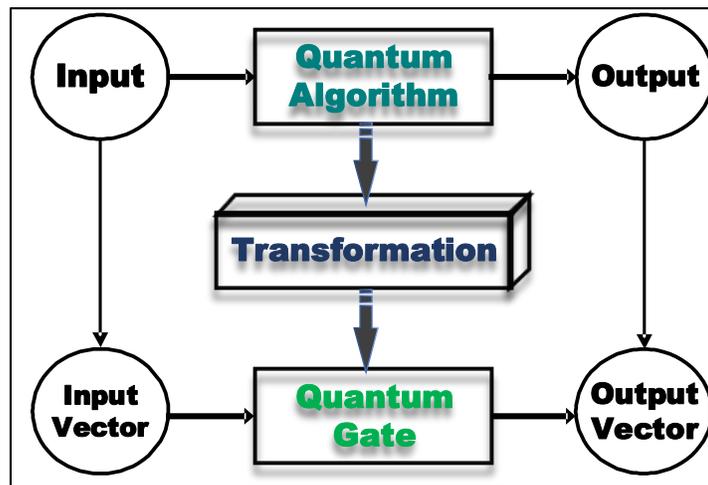

Figure 1: The gate approach for simulation of quantum algorithms using classical.

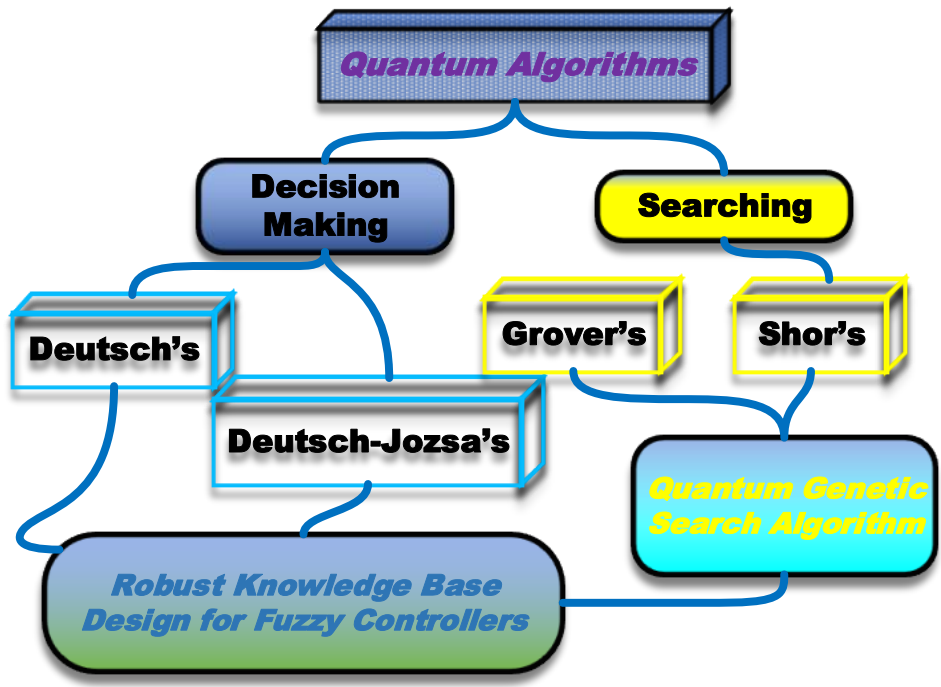

Figure 2: Classification of quantum algorithms.

In Fig. 1, an input is provided to a QA and the QA produces an output. However, the QA can be transformed to produce a quantum algorithmic gate (QAG) such that an input vector (corresponding to the QA input) is provided to the QAG to produce an output vector (corresponding to the QA output) [5].

In Fig. 2 shows classification tree of QA's for quantum soft computing and control engineering applications. QA's are either decision-making or searching as described above.

As shown, as example, in Fig. 2, Quantum Genetic Search Algorithms (QGSA) follows from Grover's and Shor's algorithms, and background for Robust KB design of Fuzzy Controllers follows from Deutsch's, Deutsch-Josa's, Grover's and/or Shor's algorithms (see, in details [4, 5]).

Let us briefly consider the design process of QAG. Figure 3 is a block diagram showing the design process of the QAG.

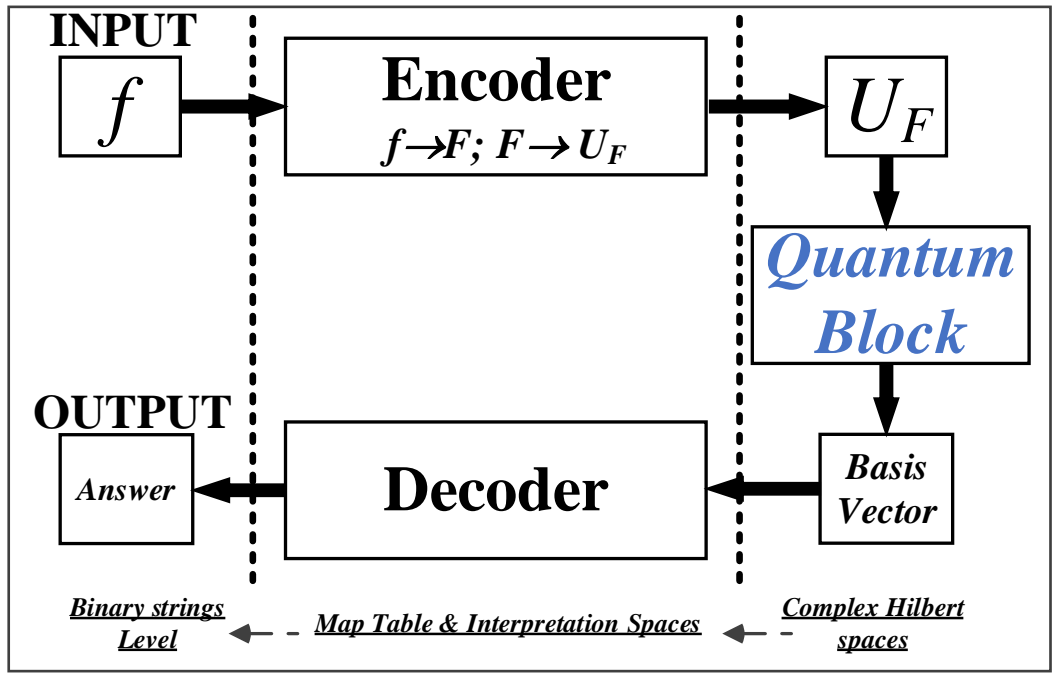

Figure 3: Schematic diagram of QA.

In Fig. 3 an input block of the QA is a function $f$ that maps binary strings into binary strings. This function $f$ is represented as a map table, defined for every string its image. The function is first encoded into a unitary matrix operator $U_F$ depending on the properties of $f$. In some sense, this operator calculates $f$ when its input and output strings are encoded into canonical basis vectors of a complex Hilbert space. The operator $U_F$ maps the vector code of every string into the vector code of its image by $f$. The quantum block operates on basis vectors in a complex Hilbert space. The vectors operated on by the quantum block are provided to a decoder, which decodes the vectors to produce an answer.

Once generated, the matrix operator $U_F$ is embedded into a quantum gate $G$. The quantum gate $G$ is a unitary matrix whose structure depends on the form of matrix $U_F$ and on the problem to be solved. The quantum gate is a unitary operator built from the dot composition of other more specific operators. The specific operators are described as tensor products of smaller matrices.

## 2 General structure of the quantum algorithmic gate (QAG) design method

Traditionally QA is written as a quantum circuit [2]. Figure 4 shows the general structure of a quantum circuit for a QAG.

As shown in Fig. 4, the general structure of the quantum circuit is based on three reversible quantum operators (superposition, entanglement, and interference) and irreversible classical operator measurement.

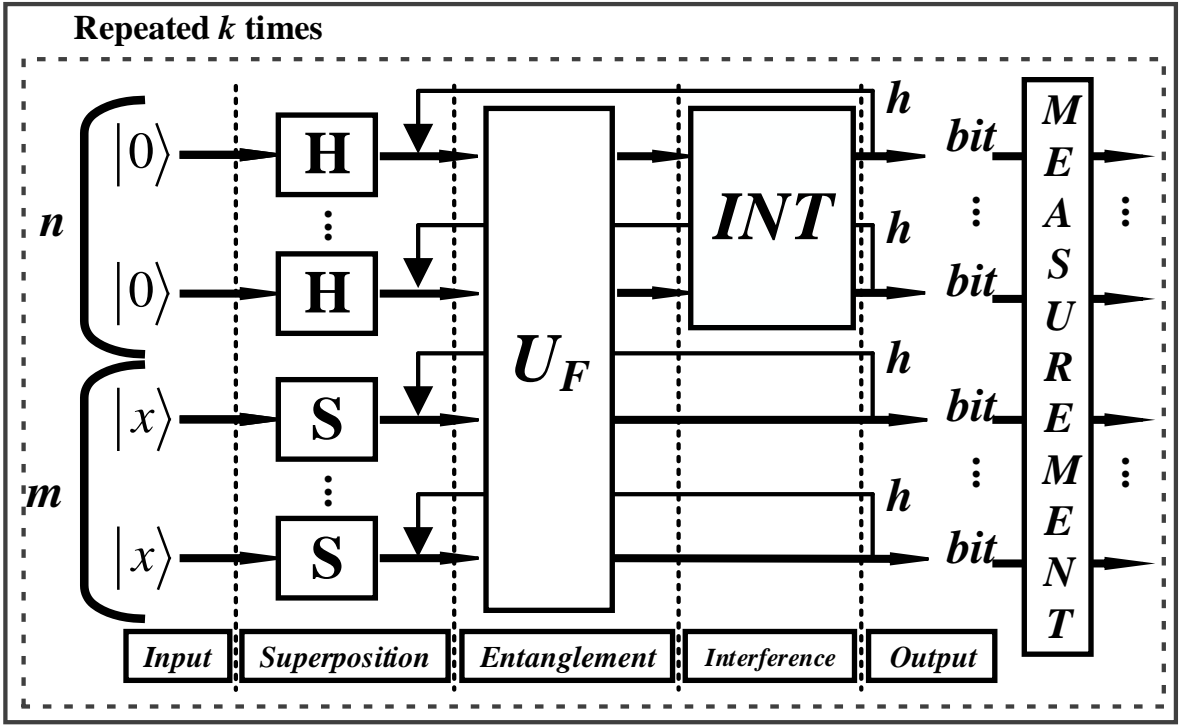

Figure 4: The structure of a quantum circuit.

The quantum circuit is a high-level description of how these smaller matrices are composed using tensor and dot products in order to generate the final quantum gate as shown in Fig. 4 (see in details [4,5]). Thus, the mathematical background of this approach is based on mappings between the quantum block operations in the complex Hilbert space [2].

The encoder and decoder operate in a map table and interpretation space, and input/output occurs on a binary string level. The Clifford and Pauli groups are the background for universal QAG design for simulation of a QA's on classical computers. Therefore, the general structure of the QAG is based on three quantum operators as superposition, entanglement, and interference, and measurement is irreversible classical operation.

The QAG acts on an initial canonical basis vector to generate a complex linear combination (called a superposition) of basis vectors as an output. This superposition contains the full information to answer the initial problem. After the superposition has been created, measurement takes place in order to extract the answer information. In quantum mechanics, a measurement is a non-deterministic operation that produces as output only one of the basis vectors in the entering superposition. The probability of every basis vector of being the output of measurement depends on its complex coefficient (probability amplitude) in the entering complex linear combination.

Thus, the segmental action of the quantum gate and of measurement makes up a quantum block (see Fig. 3). The quantum block is repeated k times in order to produce a collection of k basis vectors. Since measurement is a non-deterministic operation, these basis vectors will not necessarily be identical, and each basis vector encodes a piece of the information needed to solve the problem. The last part of the algorithm involves interpretation of the collected basis vectors in order to get the final answer for the initial problem with some probability.

## 2.1 Peculiarities of general QA - structure

A quantum algorithm (QA) calculates the qualitative properties of the function $f$. As mentioned above, QA estimates (without numerical computing) the qualitative properties of the function $f$. From a mathematical standpoint, a function $f$ is the map of one logical state into another. The problem solved by a QA can be stated in the symbolic form as follows:

*Find a certain property of function $f$ that is a map $f: \{0,1\}^n \to \{0,1\}^m$.*

The main blocks in Fig. 5 are following: i) unified operators; ii) problem-oriented operators; iii)

Benchmarks of QA simulation on classical computers; and iv) quantum control algorithms based on quantum fuzzy inference (QFI) and quantum genetic algorithm (QGA) as new types of QSA. The design process of QAG's includes the matrix design form of three quantum operators: superposition *(Sup)*, entanglement ($U_F$) - oracle, and interference *(Int)* that are the background of QA structures.

In general form, the structure of a QAG for QA in Fig. 5 can be [5] described as follows:

$$QAG = \left[ \left( Int \otimes {}^n I \right) \cdot \underbrace{\{U_F\}}_{\text{function property}} \right]^{h+1} \cdot \left[ {}^n H \otimes {}^m S \right], \quad (1)$$

where $I$ is the identity operator; the symbol $\otimes$ denotes the tensor product; $S$ is equal to $I$ or $H$ and dependent on the problem description. The heart of the quantum block is the quantum gate, which depends on the properties of matrix $U_F$. One portion of the design process in Eq. (1) is the type-choice of the entanglement problem dependent operator $U_F$ that physically describes the qualitative properties of the function $f$.

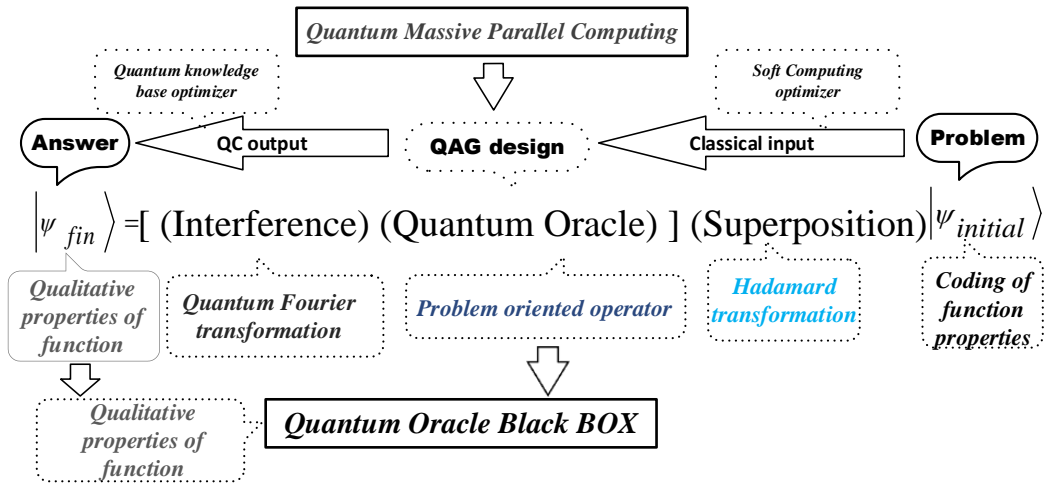

Figure 5: General structure of QA.

A general QA, written as a quantum circuit (as in Fig. 4), can be automatically translated into the corresponding programmable quantum gate for efficient classical simulation. This gate is represented as a quantum operator in matrix form such that, when it is applied to the vector input representation of the quantum register state, the result is the vector representation of the desired register output state

## 2.2 Main QAG's and main quantum operators

Three quantum operators, superposition, entanglement (quantum oracle), and interference, are the basis for quantum computations of qualitative and quantitative measures in quantum soft computing. As described above (see, Fig. 3) the structure of a QAG based on these three quantum operations of superposition, entanglement, and interference. Thus, superposition, entanglement (quantum oracle) and interference in quantum massive parallel computing are the main operators in QA.

| Algorithm | Superposition | $m$ | Interference |
|---|---|---|---|
| Deutsch's | $H \otimes I$ | 1 | $H \otimes H$ |
| Deutsch-Jozsa's | ${}^n H \otimes H$ | 1 | ${}^n H \otimes I$ |
| Grover's | ${}^n H \otimes H$ | 1 | $D_n \otimes I$ |
| Simon's | ${}^n H \otimes {}^n I$ | $n$ | ${}^n H \otimes {}^n I$ |
| Shor's | ${}^n H \otimes {}^n I$ | $n$ | $QFT_n \otimes {}^n I$ |

Table 1. Parameters of superposition and interference operators of main quantum algorithms.

The superposition operator of most QAs can be expressed as following: $Sp = \left( \bigotimes_{i=1}^{n} H \right) \otimes \left( \bigotimes_{i=1}^{m} S \right)$, where $n$ and $m$ are the numbers of inputs and of outputs respectively. Operator $S$ may be or Hadamard operator $H$ or identity operator $I$ depending on the algorithm. Numbers of outputs $n$ as well as structures of corresponding superposition and interference operators are presented in the Table 1 for different QAs on Fig. 2.

Figure 6 shows methods in QAG design. The methods as shown in Fig. 6 are based on qualitative measures of QAG design: 1) analysis of QA dynamics and structure gate design; 2) analysis of information flow; and 3) structure simulation of intelligent QA's on classical computers.

*Remark* - The analysis of information flow in [4, 5] is described. In this article analysis of QA dynamics and structure gate design, and structure simulation of intelligent QA's on classical computers are discussed.

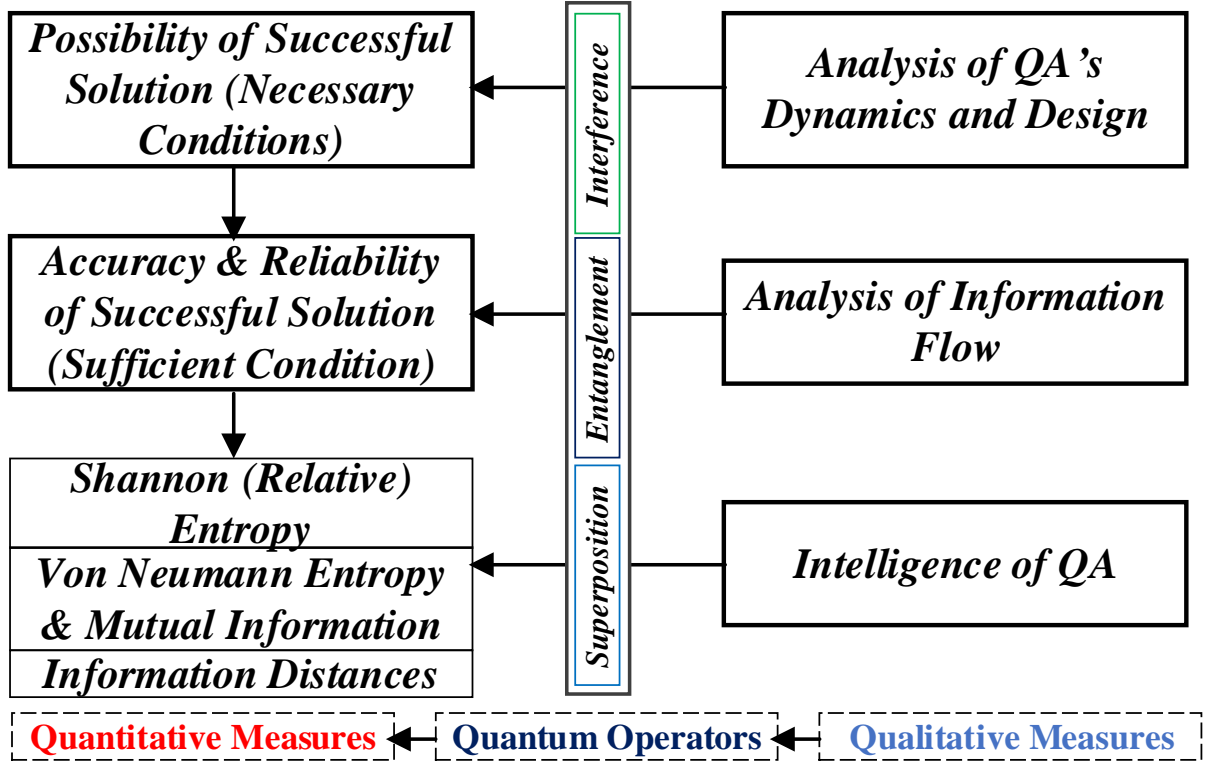

Figure 6: Methods in Quantum Algorithm Gate Design.

As shown in Fig. 6 analysis of QA dynamics provides the background for showing the existence of a solution and that the solution is unique with the desired probability. Analysis of information flow in the QA gates provides the background for showing that the unique solution exists with the desired accuracy and that the reliability of the solution can be achieved with higher probability.

The intelligence of a QA is achieved through the principle of minimum information distance between Shannon and von Neumann entropy and includes the solution of the QA stopping problem (see [5]). The output states of a QA as the solution of expected problems are the intelligent states with minimum entropic relations of uncertainty (coherent superposition states). The successful results of QA computing are robust to noise excitations in quantum gates, and intelligent quantum operations are fault-tolerant in quantum soft computing [5].

With the method of quantum gate design presented herein, various different structures of QA can be realized (see, Fig. 4), as shown in Table 2 below.

Remark. A quantum computer is difficult to build because of decoherence effects. Decoherence introduces errors in the superposition. The decoherence problem is reduced by using tools of quantum soft computing such as a quantum genetic search algorithm (QGSA). Errors produced by decoherence are of three kinds: (i) phase errors; (ii) bit-flip errors; and (iii) both phase and bit-flip errors.

These three errors can all be modeled using unitary transformations [5].

This means that if the QGSA is implemented on a physical quantum-mechanical system, one would gain the advantages of quantum parallelism and reduce the problem of decoherence, because decoherence can be used as a natural generator of mutation and crossover operators.

Let us discuss briefly any mathematical backgrounds and its physical peculiarities for quantum computing based on QAG.

| Name | Algorithm | Gate Symbolic Form: $\left[ \underbrace{(Int \otimes {}^m I)}_{Interference} \cdot \underbrace{U_F}_{Entanglement} \right]^{h+1} \cdot \underbrace{({}^n H \otimes {}^m S)}_{Superposition}$ |
|---|---|---|
| Deutsch-Jozsa (D. – J.) | $m=1;\ S=H\ (x=1)$<br>$Int = {}^n H$ ;<br>$k=1\ h=0$ | $({}^n H \otimes I) \cdot U_F^{D-J} \cdot ({}^{n+1} H)$ |
| Simon (Sim) | $m=n,\ S=I$<br>$(x=0)\ Int = {}^n H\ k=O(n)$<br>$h=0$ | $({}^n H \otimes {}^n I) \cdot U_F^{Sim} \cdot ({}^n H \otimes {}^n I)$ |
| Shor (Shr) | $m=n$<br>$S=I\ (x=0)$<br>$Int = QFT_n$<br>$k=O(Poly(n))\ h=0$ | $(QFT_n \otimes {}^n I) \cdot U_F^{Shr} \cdot ({}^n H \otimes {}^n I)$ |
| Grover (Gr) | $m=1$<br>$S=H\ (x=1)$<br>$Int = D_n$<br>$k=1$<br>$h=O(2^{n/2})$ | $(D_n \otimes I) \cdot U_F^{Gr} \cdot ({}^{n+1} H)$ |

Table 2. Quantum gate parameters for QA's structure design.

## 2.3 Design technology of quantum algorithmic gates and simulation system

The searching problem can be stated in terms of a list $\mathcal{L}[0,1,\ldots,N-1]$ with a number $N$ of unsorted elements. Denote by $x_0$ the marked element in $\mathcal{L}$ that are sought. The quantum mechanical solution of this searching problem goes through the preparation of a quantum register in a quantum computer to store the $N$ items of the list. This will allow exploiting quantum parallelism. Thus, assume that the quantum registers are made of $n$ source qubits so that $N = 2^n$.

A target qubit is used to store the output of function evaluations or calls. To implement the quantum search, construct a unitary operation that discriminates between the marked item $x_0$ and the rest. The following function:

$$f_{x_0}(x) = \begin{cases} 0, & \text{if } x \neq x_0 \\ 1, & \text{if } x = x_0 \end{cases}, \qquad (2)$$

and its corresponding unitary operation $U_{f_{x_0}}|x\rangle|y\rangle = |x\rangle|y \oplus f_{x_0}(x)\rangle$. It is assumed the access to $f$ via the following quantum oracle: $U_f|0,0\rangle = |0,f(0)\rangle,\ U_f|1,0\rangle = |0,f(1)\rangle$. After these two queries, we can measure qubit 1 with a deterministic outcome, and answer whether $f(0) = f(1)$. However, a quantum checker can apply $U_f$ to a linear combination of states in the computational basis. Count how many applications of this operation or oracle calls are needed to find the item. The rationale behind the Grover algorithm is: 1) to start with a quantum register in a state where the computational basis states are equally present; 2) to apply several unitary transformations to produce an outcome state in which the probability of catching the marked state $|x_0\rangle$ is large enough.

The steps in Grover's algorithm are shown in tabular form below (the quantum circuit shown in Fig. 13 (a).)

| Steps | Computational algorithm | Formula |
|---|---|---|
| *Step 1* | Initialize the quantum registers to the state: $$|\psi_1 = input\rangle := |00\ldots0\rangle|1\rangle$$ | (3) |
| *Step 2* | Apply bit-wise the Hadamard one-qubit gate to the source register, so as to produce a uniform superposition of basis states in the source register, and also to the target register: $$|\psi_2\rangle := U_H^{\otimes(n+1)}|\psi_1\rangle = \frac{1}{2^{(n+1)/2}} \sum_{x=0}^{2^n-1} |x\rangle \sum_{y=0,1} (-1)^y |y\rangle.$$ | (4) |
| *Step 3* | Apply the operator $U_{f_{x_0}}$: $$|\psi_3\rangle := U_{f_{x_0}}|\psi_2\rangle = \frac{1}{2^{(n+1)/2}} \sum_{x=0}^{2^n-1} (-1)^{f_{x_0}(x)} |x\rangle \sum_{y=0,1} (-1)^y |y\rangle.$$ Let $U_{x_0}$ be the operator by $$U_{x_0}|x\rangle := (1 - 2|x_0\rangle\langle x_0|)|x\rangle = \begin{cases} |x\rangle, & \text{if } x \neq x_0 \\ -|x_0\rangle, & \text{if } x = x_0 \end{cases},$$ that is, it flips the amplitude of the marked state leaving the remaining source basis states unchanged. The state in the source register of Step 3 equals precisely the result of the action of $U_{x_0}$, i.e., $|\psi_3\rangle = \left((1-2|x_0\rangle\langle x_0|)\otimes 1\right)|\psi_2\rangle$. | (5) |
| *Step 4* | Apply next the operation $D$ known as inversion about the average. This operator is defined as follows $$D := -\left(U_H^{\otimes n} \otimes I\right) U_{f_0} \left(U_H^{\otimes n} \otimes I\right),$$ and $$|output\rangle = D|\psi_3\rangle$$ where $U_{f_0}$ is the operator in Step 3 for $x_0 = 0$. The effect of this operator on the source is to transform $$\sum_x \alpha_x |x\rangle \mapsto \sum_x \left(-\alpha_x + \langle\alpha\rangle|x\rangle\right),$$ where $\langle\alpha\rangle := 2^{-n}\sum_x \alpha_x$ is the mean of the amplitudes, so its net effect is to amplify the amplitude of $|x_0\rangle$ over the rest. | (6) |
| *Step 5* | Iterate Steps 3 and 4 a number of times $m$. | |
| *Step 6* | Measure the source qubits (in the computational basis). The number $m$ is determined such that the probability of finding the searched item $x_0$ is maximal. | |

According to Steps 2 - 4 above and (1), the QAG of Grover's quantum search algorithm (QSA) is $G = (D_n \otimes I) \cdot U_F \cdot \left(^n H \otimes H\right)$ that acts on the initial state of both registers in the QSA.

Computational analysis of Grover's QSA is similar to analysis of the Deutsch-Jozsa QA. The basic component of the algorithm is the quantum operation encoded in Steps 3 and 4, which is repeatedly applied to the uniform state $|\psi_2\rangle$ in order to find the marked element. Steps 5 and 6 in Grover's algorithm are also applied in Shor's QSA. Although this procedure resembles the classical strategy, Grover's operation enhances by constructive interference of quantum amplitudes the presence of the marked state.

# 3 Computational models of quantum search algorithms

We have considered in [4] *five* practical approaches to design fast algorithms for the simulation most of known QAs on classical computers:

1. *Matrix based approach*;
2. *Model representations* of quantum operators in fast QAs;
3. *Algorithmic based approach,* when matrix elements are calculated on "demand";
4. *Problem-oriented approach,* where we succeeded to run Grover's algorithm with up to 64 and more qubits with Shannon entropy calculation (up to 1024 without termination condition);
5. *Quantum algorithms with reduced number of operators* (entanglement-free QA, and so on).

Detail description of these approaches is given in [4].

Figure 7 shows the structure description of the QA Benchmark Block.

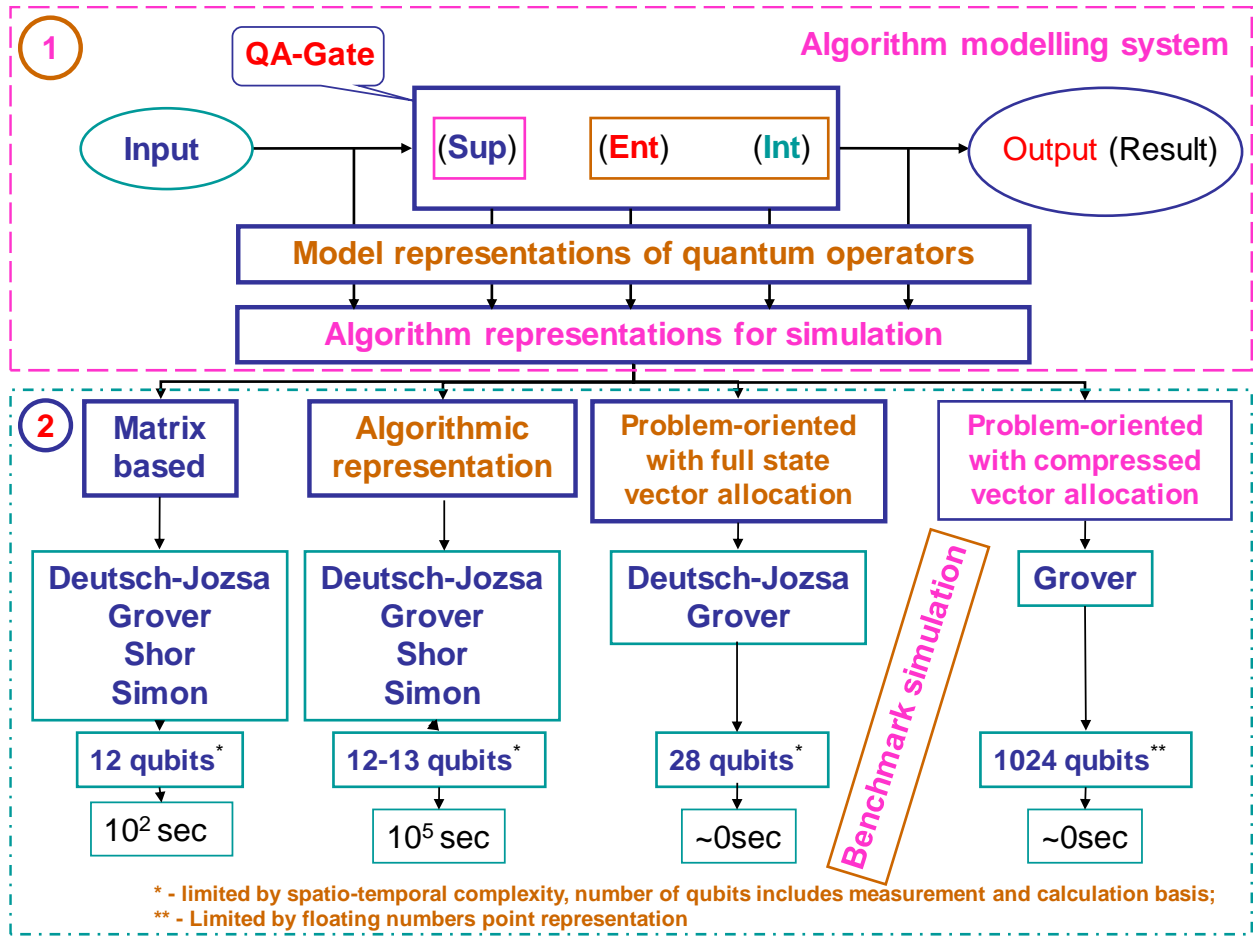

Figure 7: Algorithm modeling system in QFMS.

The efficient implementations of a number of operations for quantum computation include controlled phase adjustment of the amplitudes in the superposition, permutation, approximation of transformations and generalizations of the phase adjustments to block matrix transformations. These operations generalize those used as example in quantum search algorithms (QSA's) that can be realized on a classical computer. The application of this approach is applied herein to the efficient simulation on classical computers of the Deutsch QA, the Deutsch–Jozsa QA, the Simon QA, the Shor QA and the Grover QA.

Implementation of a QA is based on a QAG. In the language of classical computing, a quantum computer is programmed by designing a QAG. The prior art reports relatively few such gates because the basic principles underlying the quantum version of programming are in their infancy and algorithms to date have been programmed by ad-hoc techniques.

The problems solved by the QA can be stated (1) as follows:

| Input | A function $f: \{0,1\}^n \to \{0,1\}^m$ |
|---|---|
| **Problem** | Find a certain property of $f$ |

The structure of a quantum operator $U_F$ in QA's as shown in block of Fig. 3 is outlined, with a high-level representation, in the scheme diagram Fig. 1. In Fig. 3 the input of the QA is a function $f$ that maps from binary strings into binary strings. This function is represented as a map table, defining for every string its image. The function $f$ is encoded according to an $F$ - truth table. The function is transformed according to a transform $U_F$ - truth table into a unitary matrix operator $U_F$ depending on $f$'s properties. In some sense, this operator calculates $f$ when its input and output strings are encoded into canonical basis vectors of a complex Hilbert space: $U_F$ maps the vector code of every string into the vector code of its image by $f$. A squared matrix $U_F$ on the complex field is unitary *if and only if (iff)* its inverse matrix coincides with its conjugate transpose: $U_F^{-1} = U_F$. A unitary matrix is always reversible and preserves the norm of vectors.

Figure 8 shows structure of the quantum block from Fig. 3.

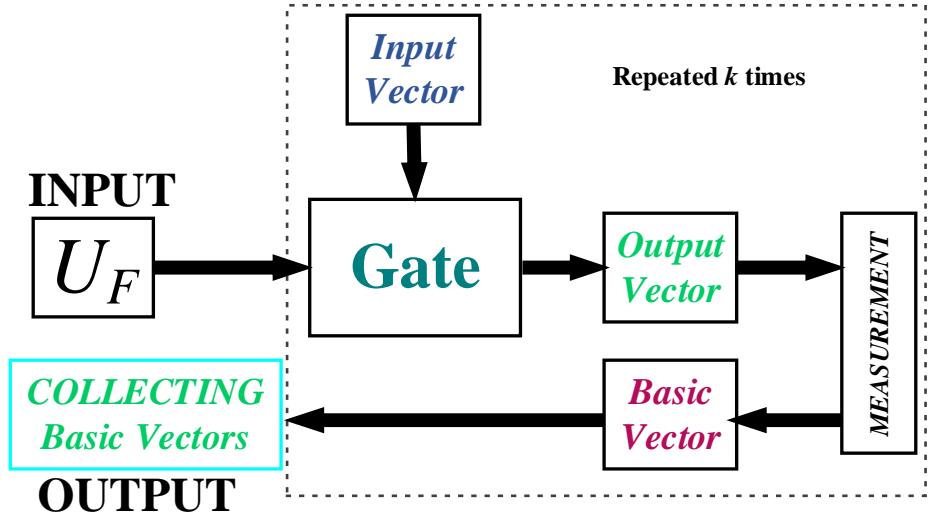

Figure 8: Structure of Quantum Block in Fig. 3.

In the structure, the matrix operator $U_F$ has been generated it is embedded into a quantum gate as a QAG, a unitary matrix whose structure depends on the form of matrix $U_F$ and on the problem to be solved. In the QA, the QAG acts on an initial canonical basis vector (which can always choose the same vector) in order to generate a complex linear combination (superposition) of basis vectors as output. This superposition contains all the information to answer the initial problem.

After this superposition has been created, in measurement block takes place in order to extract this information. In quantum mechanics, measurement is a non-deterministic operation that produces as output only one of the basis vectors in the entering superposition. The probability of every basis vector of being the output of measurement depends on its complex coefficient (probability amplitude) in the entering complex linear combination.

The segmental action of the QAG and of measurement characterizes the quantum block in Fig. 8. The quantum block is repeated $k$ times in order to produce a collection of $k$ basis vectors. Since measurement a nondeterministic operation, these basic vectors are not be necessarily identical and each one of them will encode a piece of the information needed to solve the problem. The collection block in Fig. 8 of the algorithm outputs the interpretation of the collected basis vectors in order to get the answer for the initial problem with a certain probability.

## 3.1 Encoder

The behavior of the encoder in Fig. 3 is described in the scheme diagram of Fig. 9.

Function $f$ is encoded into matrix $U_F$ in three steps.

In *step* 1, the map table ($f$ − truth table) of function $f$: $\{0,1\}^n \to \{0,1\}^m$ is transformed into the map table ($F$ − truth table) of the injective function $F$: $\{0,1\}^{n+m} \to \{0,1\}^{n+m}$ such that:

$$F(x_0, .., x_{n-1}, y_0, .., y_{m-1}) = (x_0, .., x_{n-1}, f(x_0, .., x_{n-1}) \oplus (y_0, .., y_{m-1})).$$

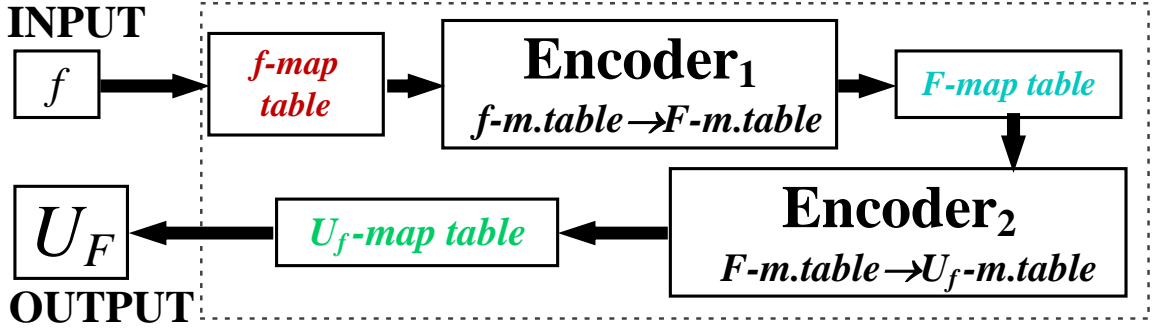

Figure 9: The encoder block scheme diagram.

*Remark* - The need to deal with an injective function comes from the requirement that $U_F$ is unitary. A unitary operator is reversible, so it cannot map two different inputs in the same output. Since $U_F$ will be the matrix representation of $F$, $F$ is injective. If one directly employed the matrix representation of function $f$, one could obtain a non-unitary matrix, since $f$ could be non-injective. So, injectivity is fulfilled by increasing the number of bits and considering function $F$ instead of function $f$. The function $f$ can be calculated from $F$ by putting $(y_0,...,y_{m-1}) = (0,...,0)$ in the input string and reading the last $m$ values of the output string.

Reversible circuits realize permutation operations. It is possible to realize any Boolean circuit $F: \mathbb{B}^n \to \mathbb{B}^m$ by reversible circuit. For this case, one need not calculate the function $F: \mathbb{B}^n \to \mathbb{B}^m$. One can calculate another function with expanding $F_\oplus : \mathbb{B}^{n+m} \to \mathbb{B}^{n+m}$ that is defined as following relation: $F_\oplus(x, y) = (x, y \oplus F(x))$ where the operation $\oplus$ is defined as addition on module 2.

Then the value of $F(x)$ is defined as $F_\oplus(x,0) = (x, F(x))$. For example, the *XOR* operator between two binary strings $p$ and $q$ of length $m$ is a string $s$ of length $m$ such that the $i$-th digit of $s$ is calculated as the exclusive *OR* between the $i$-th digits of $p$ and $q$:

$$p = (p_0, …, p_{n-1}), q = (q_0, …, q_{n-1}); s = p \oplus q = ((p_0 + q_0) \bmod 2, …, (p_{n-1} + q_{n-1}) \bmod 2)).$$

In *step* 2, the function from $F$ map table is transformed into $U_F$ map table, according to the following constraint:

$$\forall s \in \{0,1\}^{n+m}: U_F[\tau(s)] = \tau[F(s)] \tag{7}$$

The code map $\tau: \{0,1\}^{n+m} \to \mathbb{C}^{2^{n+m}}$ ($\mathbb{C}^{2^{n+m}}$ is the target Complex Hilbert Space) is such that:

$$\tau(0) = \begin{pmatrix} 1 \\ 0 \end{pmatrix} = |0\rangle, \ \tau(1) = \begin{pmatrix} 0 \\ 1 \end{pmatrix} = |1\rangle; \ \tau(x_0,…,x_{n+m-1}) = \tau(x_0) \otimes … \otimes \tau(x_{n+m-1}) = |x_0 … x_{n+m-1}\rangle$$

Code $\tau$ maps bit values into complex vectors of dimension 2 belonging to the canonical basis of $\mathbb{C}^2$. Besides, using tensor product, $\tau$ maps the general state of a binary string of dimension $n$ into a vector of dimension $2^n$, reducing this state to the joint state of the $n$ bits composing the register. Every bit state is transformed into the corresponding 2-dimesional basis vector and then the string state is mapped into the corresponding $2^n$-dimesional basis vector by composing all bit-vectors through tensor product. In this sense tensor product is the vector counterpart of state conjunction. Basis vectors are denoted using the *ket* notation $|i\rangle$. This notation is taken from Dirac description of quantum mechanics.

In *step* 3, the $U_F$ map table is transformed into $U_F$ using the following transformation rule:

$$[U_F]_{ij} = 1 \Leftrightarrow U_F|j\rangle = |i\rangle.$$

This rule can be understood by considering vectors $|i\rangle$ and $|j\rangle$ as column vectors. These vectors belong to the canonical basis, where $U_F$ defines a permutation map of the identity matrix rows. In general, row $|j\rangle$ is mapped into row $|i\rangle$.

## 3.2 Quantum block

The heart of the quantum block is the quantum gate, which depends on the properties of matrix $U_F$. The quantum block uses the QAG, which depends on the properties of matrix $U_F$. The structure of a quantum operator $U_F$ in QA's as shown in Fig. 3 is outlined, with a high-level representation, in the scheme diagram of Fig. 8.

The scheme in Fig. 8 gives a more detailed description of the quantum block. The matrix operator $U_F$ of Fig. 9 is the output of the encoder block represented in Fig. 3.

Here, it becomes the input for the quantum block. This matrix operator is embedded into a more complex gate: the gate $G$ (QAG). Unitary matrix $G$ is applied $k$ times to an initial canonical basis vector $|i\rangle$ of dimension $2^{n+m}$. Each time, the resulting complex superposition $G|0\ldots01\ldots1\rangle$ of basis vectors is measured in measurement block, producing one basis vector $|x_i\rangle$ as result. The measured basis vectors $\{x_1,\ldots,x_k\}$ are collected together in block of basis vectors.

This collection is the output of the quantum block. The "intelligence" of the QA's is in the ability to build a QAG that is able to extract the information necessary to find the required property of *f* and to store it into the output vector collection.

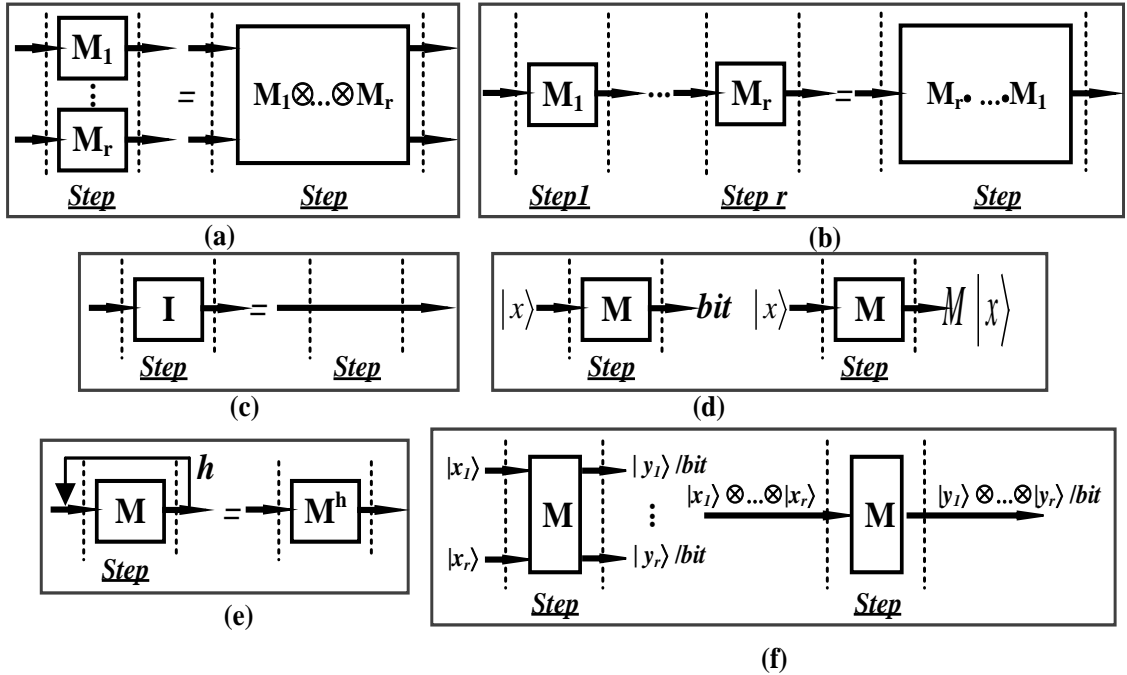

Figure 10: Transformation rules.

In order to represent QAGs it is useful to employ some diagrams called quantum circuits, as shown in Fig. 4. Each rectangle is associated with a matrix $2^n \times 2^n$, where $n$ is the number of lines entering and leaving the rectangle. For example, the rectangle marked $U_F$ is associated with the matrix $U_F$.

Using a high-level description of the gate and, using transformation rules shown in Fig. 10, it is possible to compile the corresponding gate-matrix.

*Remark.* These rules are listed in Fig. 10 as following: (a) Rule 1 — Tensor Product Transformation; (b) Rule 2 — Dot Product Transformation; (c) Rule 3 — Identity Transformation; (d) Rule 4 — Propagation Rule; (e) Rule 5 — Iteration Rule; and (f) Rule 6 — Input / Output Tensor Rule. It will be clearer how to use these rules when we afford the first examples of quantum algorithm.

### 3.3 Decoder

The decoder block of Fig. 3 interprets the basis vectors (collected in block basis vectors) of after the iterated execution in the quantum block. Decoding these vectors involves retranslating them into binary strings and interpreting them directly in decoder block if they already contain the answer or use them, for instance as coefficients vectors for some equation system, in order to get the searched solution.

## 4 Grover's Problem statement

Grover's quantum searching problem is stated as following:

| Input | Given a function $f:\{0,1\}^n \to \{0,1\}$ such that $\exists\, x \in \{0,1\}^n: (f(x) = 1 \land \forall\, y \in \{0,1\}^n: x \neq y \Rightarrow f(y) = 0)$ |
|---|---|
| Problem | Find $x$ |

Figure 11 shows the definition of the Grover's problem.

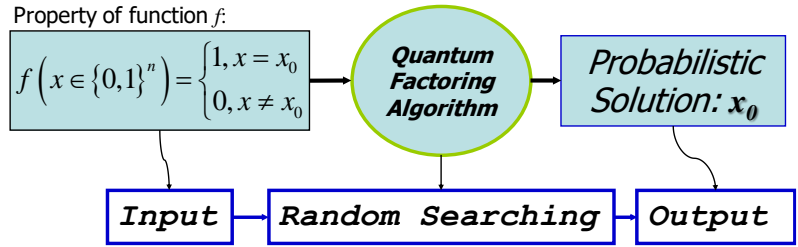

Figure 11: Grover's QA: Problem definition.

Figure 12 shows step design definitions in Grover's QA.

| N | Design step definition |
|---|---|
| 0 | **Step 0: Encoder** <br> Step 0.1: Injective function F building <br> Step 0.2: Preparation of map table for entanglement operator $U_F$ |
| 1 | **Step 1: Preparation of quantum operators** <br> Step 1.1: Preparation of superposition operator <br> Step 1.2: Preparation of entanglement operator using information from step 0.2 <br> Step 1.3: Preparation of interference operator <br> Step 1.4: Quantum gate assembly |
| 2 | **Step 2: Algorithm execution** <br> Step 2.1: Application of superposition operator <br> Step 2.2: Application of entanglement operator <br> Step 2.3: Application of interference operator <br> Step 2.4: Repeat steps 2.2 and 2.3 **h** times <br> Step 2.5: Measurement and interpretation of the output |

Figure 12: Grover's QA: Steps of the algorithm design.

*Encoder* - In order to make the discussion more comprehensible, it is convenient to first consider a special function with n = 2, then the general case with n = 2 is discussed, and finally to analyze the general case with n > 0.

## 4.1 Design process of Grover's QAG

Let us consider the implementation of Grover QSA steps in QAG design.

**A.** *Introductory example* Consider the case: $n = 2, f(01) = 1$. In this case the $f$ map table (see, Fig. 9) is defined by:

| $x$ | $f(x)$ |
|---|---|
| 00 | 0 |
| 01 | 1 |
| 10 | 0 |
| 11 | 0 |

*Step 1*

Function $f$ is encoded into injective function $F$, built according to the usual statement:

$$F : \{0,1\}^{n+1} \to \{0,1\}^{n+1}; F(x_0, x_1, y_0) = (x_0, x_1, f(x_0, x_1) \oplus y_0)$$

Then the $F$ map table is:

| $(x_0, x_1, y_0)$ | $F(x_0, x_1, y_0)$ |
|---|---|
| 000 | 000 |
| 010 | 011 |
| 100 | 100 |
| 110 | 110 |
| 001 | 001 |
| 011 | 011 |
| 101 | 101 |
| 111 | 111 |

*Step 2*

Now encode $F$ into the map table of $U_F$ using the usual rule: $\forall s \in \{0,1\}^{n+1}: U_F[\tau(s)] = \tau[F(s)]$ where $\tau$ is the code map defined in above. This means:

| $\lvert x_0 x_1 y_0 \rangle$ | $U_F \lvert x_0 x_1 y_0 \rangle$ |
|---|---|
| $\lvert 000 \rangle$ | $\lvert 000 \rangle$ |
| $\lvert 010 \rangle$ | $\lvert 011 \rangle$ |
| $\lvert 100 \rangle$ | $\lvert 100 \rangle$ |
| $\lvert 110 \rangle$ | $\lvert 110 \rangle$ |
| $\lvert 001 \rangle$ | $\lvert 001 \rangle$ |
| $\lvert 011 \rangle$ | $\lvert 011 \rangle$ |
| $\lvert 101 \rangle$ | $\lvert 101 \rangle$ |
| $\lvert 111 \rangle$ | $\lvert 111 \rangle$ |

*Step 3*

From the map table of $U_F$ calculate the corresponding matrix operator. This matrix is obtained using the rule: $[U_F]_{ij} = 1 \Leftrightarrow U_F \lvert j \rangle = \lvert i \rangle$. $U_F$ is thus:

| $U_F$ | $\lvert 00 \rangle$ | $\lvert 01 \rangle$ | $\lvert 10 \rangle$ | $\lvert 11 \rangle$ |
|---|---|---|---|---|
| $\lvert 00 \rangle$ | I | 0 | 0 | 0 |
| $\lvert 01 \rangle$ | 0 | C | 0 | 0 |
| $\lvert 10 \rangle$ | 0 | 0 | I | 0 |
| $\lvert 11 \rangle$ | 0 | 0 | 0 | I |

The effect of this matrix is to leave unchanged the first and the second input basis vectors of the input tensor product, flipping the third one when the first vector is $\lvert 0 \rangle$ and the second is $\lvert 1 \rangle$. This agrees with the constraints on $U_F$ stated above.

**B.** *General case with $n = 2$.* Now take into consideration the more general case: $n = 2, f(\underline{x}) = 1$. The corresponding matrix operator is:

| $U_F$ | $\lvert 00 \rangle$ | $\lvert 01 \rangle$ | $\lvert 10 \rangle$ | $\lvert 11 \rangle$ |
|---|---|---|---|---|
| $\lvert 00 \rangle$ | $M_{00}$ | 0 | 0 | 0 |
| $\lvert 01 \rangle$ | 0 | $M_{01}$ | 0 | 0 |
| $\lvert 10 \rangle$ | 0 | 0 | $M_{10}$ | 0 |
| $\lvert 11 \rangle$ | 0 | 0 | 0 | $M_{11}$ |

with $M_{\underline{x}} = C \land \forall i \neq \underline{x}: M_i = I$.

**C. General case** It is relatively simple now to generalize operator $U_F$ from the case $n = 2$ to the case $n > 1$. The operator $C$ is found on the main diagonal of the block matrix, in correspondence of the celled labeled by vector $|\underline{x}\rangle$, where $\underline{x}$ is the binary string having image one by $f$. Therefore:

| $U_F$ | $|00\rangle$ | $|01\rangle$ | ... | $|11\rangle$ |
|---|---|---|---|---|
| $|00\rangle$ | $M_{00}$ | 0 | ... | 0 |
| $|01\rangle$ | 0 | $M_{01}$ | ... | 0 |
| ... | ... | ... | ... | ... |
| $|11\rangle$ | 0 | 0 | ... | $M_{11}$ |

with $M_{\underline{x}} = C \land \forall i \neq \underline{x}: M_i = I$.

*Quantum block*

The matrix $U_F$, the output of the encoder, is embedded into the QAG.
This gate is described in Fig. 13, a, using a quantum circuit of Grover QSA.

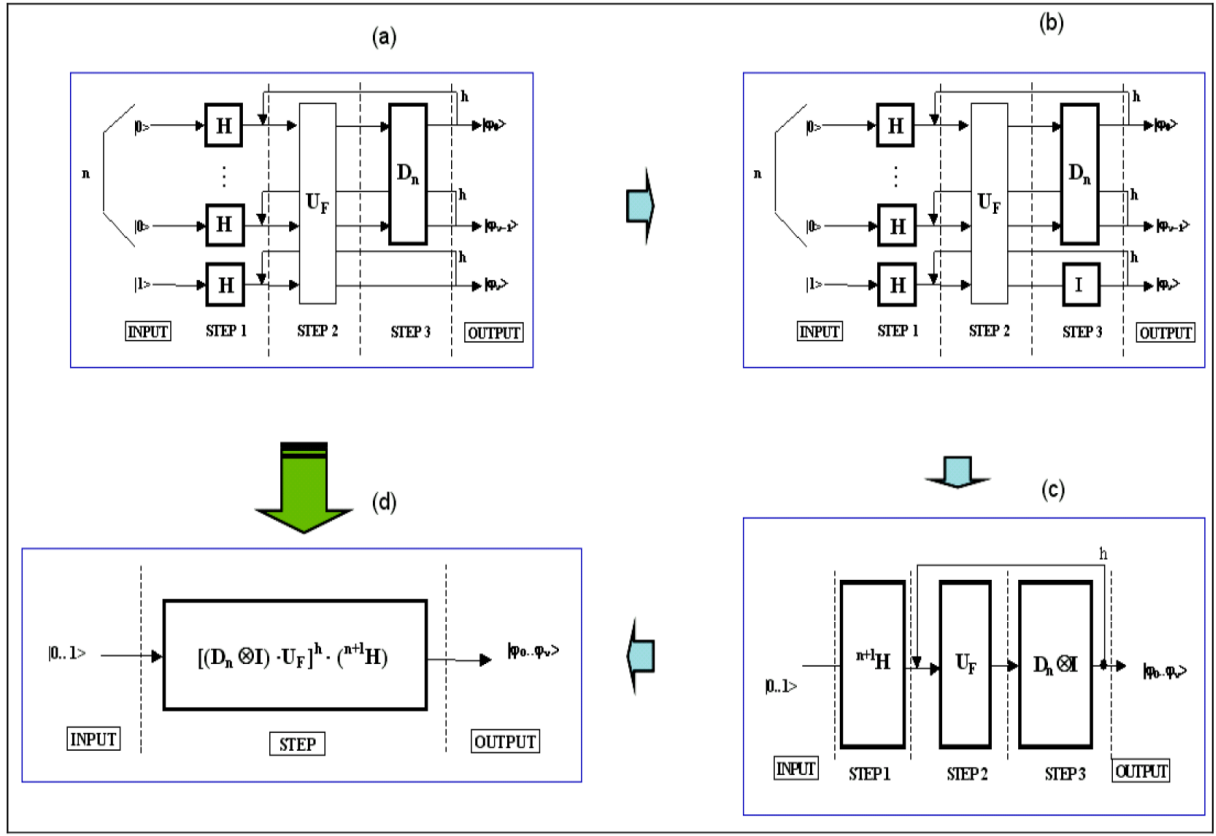

Figure 13: Grover's quantum algorithm simulation: Circuit representation and corresponding gate design.

Operator $D_n$ is called a "*diffusion matrix*" of order $n$ and it is responsible for interference in this algorithm. It plays the same role as the $QFT_n$ in Shor's algorithm and of $^nH$ in Deutsch-Jozsa's and Simon's algorithms. This matrix is defined as:

| $D_n$ | $|0..0\rangle$ | $|0..1\rangle$ | ... | $|i\rangle$ | ... | $|1..0\rangle$ | $|1..1\rangle$ |
|---|---|---|---|---|---|---|---|
| $|0..0\rangle$ | $-1+1/2^{n-1}$ | $1/2^{n-1}$ | ... | $1/2^{n-1}$ | ... | $1/2^{n-1}$ | $1/2^{n-1}$ |
| $|0..1\rangle$ | $1/2^{n-1}$ | $-1+1/2^{n-1}$ | ... | $1/2^{n-1}$ | ... | $1/2^{n-1}$ | $1/2^{n-1}$ |
| ... | ... | ... | ... | ... | ... | ... | ... |
| $|i\rangle$ | $1/2^{n-1}$ | $1/2^{n-1}$ | ... | $-1+1/2^{n-1}$ | ... | $1/2^{n-1}$ | $1/2^{n-1}$ |
| ... | ... | ... | ... | ... | ... | ... | ... |
| $|1..0\rangle$ | $1/2^{n-1}$ | $1/2^{n-1}$ | ... | $1/2^{n-1}$ | ... | $-1+1/2^{n-1}$ | $1/2^{n-1}$ |
| $|1..1\rangle$ | $1/2^{n-1}$ | $1/2^{n-1}$ | ... | $1/2^{n-1}$ | ... | $1/2^{n-1}$ | $-1+1/2^{n-1}$ |

Using Rule 3 from Fig. 2.4, compile the previous circuit into the circuit presented as in the Fig. 13, b, and then into the circuit of Fig. 13, c and using rule 2 in Fig. 9 design on Fig. 13, d.

## 4.2 Computer design process of Grover's QAG (Gr-QAG) and simulation results

Consider the design process of Grover's QAG according to the steps represented in Fig. 12. Figure 14 shows Step 0, the encoding process, for the case of order $n = 3$ and answer search 1.

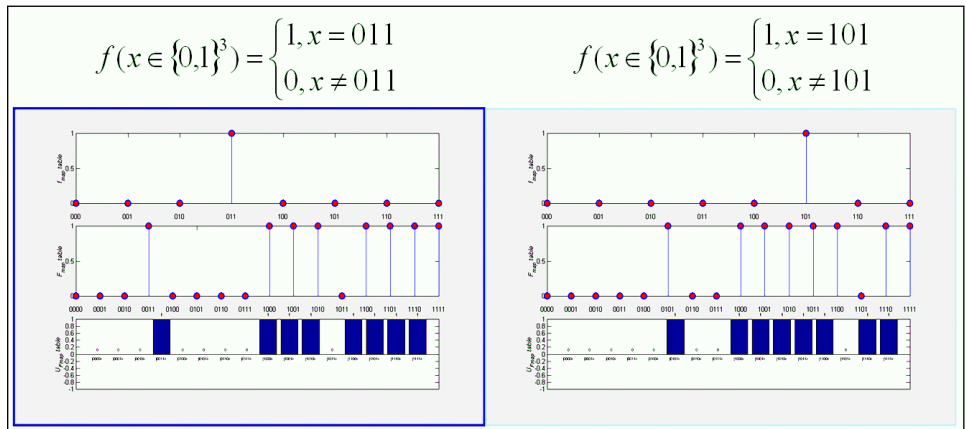

Figure 14: Grover's QA: Step 0. Encoding: Order n =3, 1 answer search.

For comparison the similar results for the cases of order $n = 3$ and answer search 2 and 3 are shown in Fig. 15.

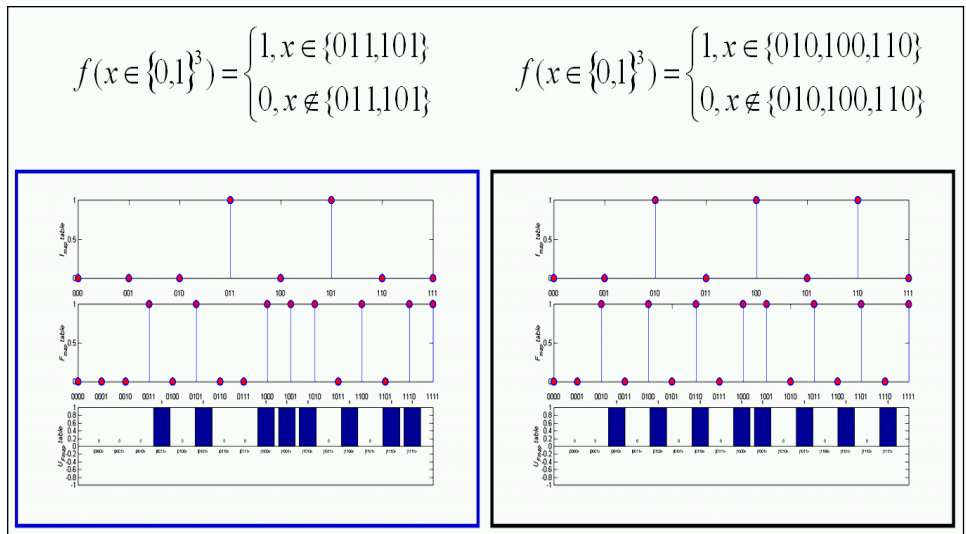

Figure 15: Grover's QA: Step 0. Encoding: Order n = 3, 2 and 3 answers search.

Figure 16 shows Step 1.1 (from Fig. 12) for design of the superposition operator.

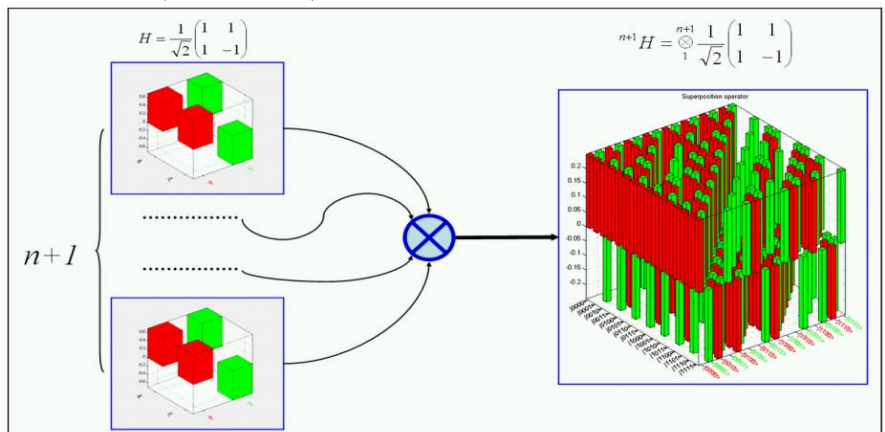

Figure 16: Grover's QA: Step 1.1. Preparation of quantum operators: Superposition operator.

Preparation of quantum entanglement (step 1.2 from Fig. 12) for the one answer search is shown in Fig. 17.

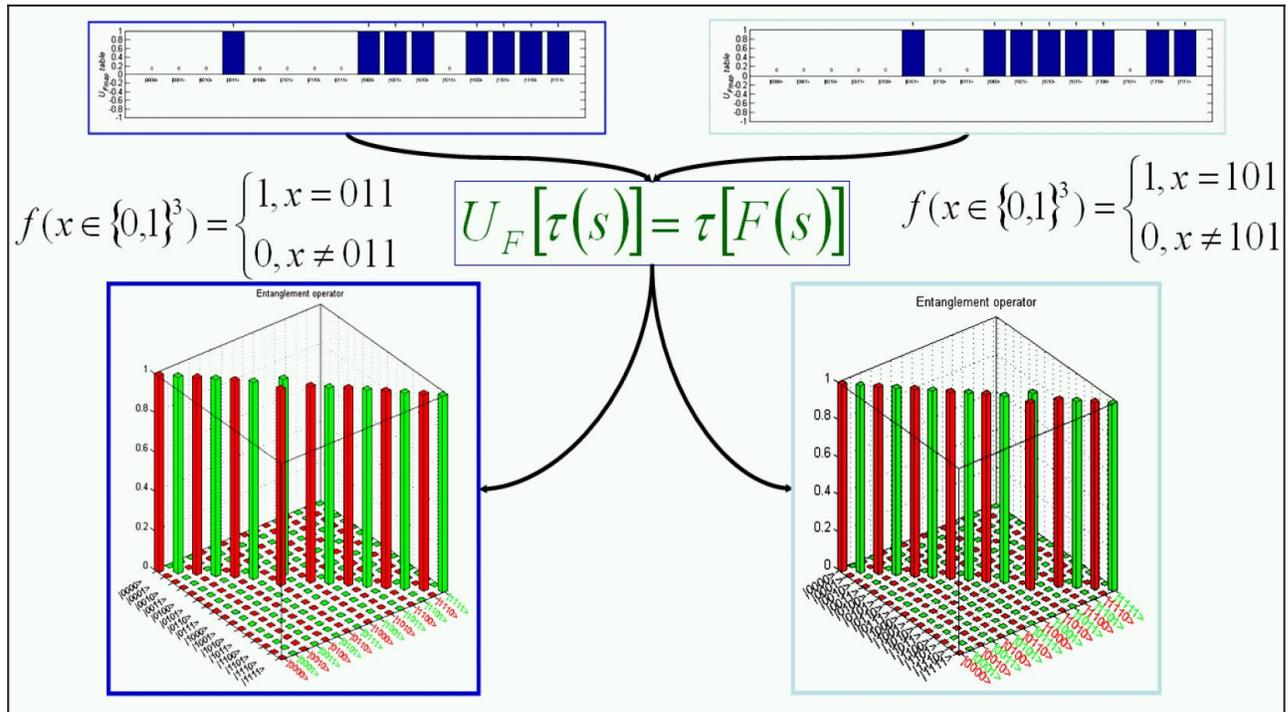

Figure 17: Grover's QA: Step 1.2. Preparation of quantum operators: Entanglement operators for 1 answer search.

The cases for 2 and 3 answer searches if the preparation of the entanglement operator is shown in Fig. 18.

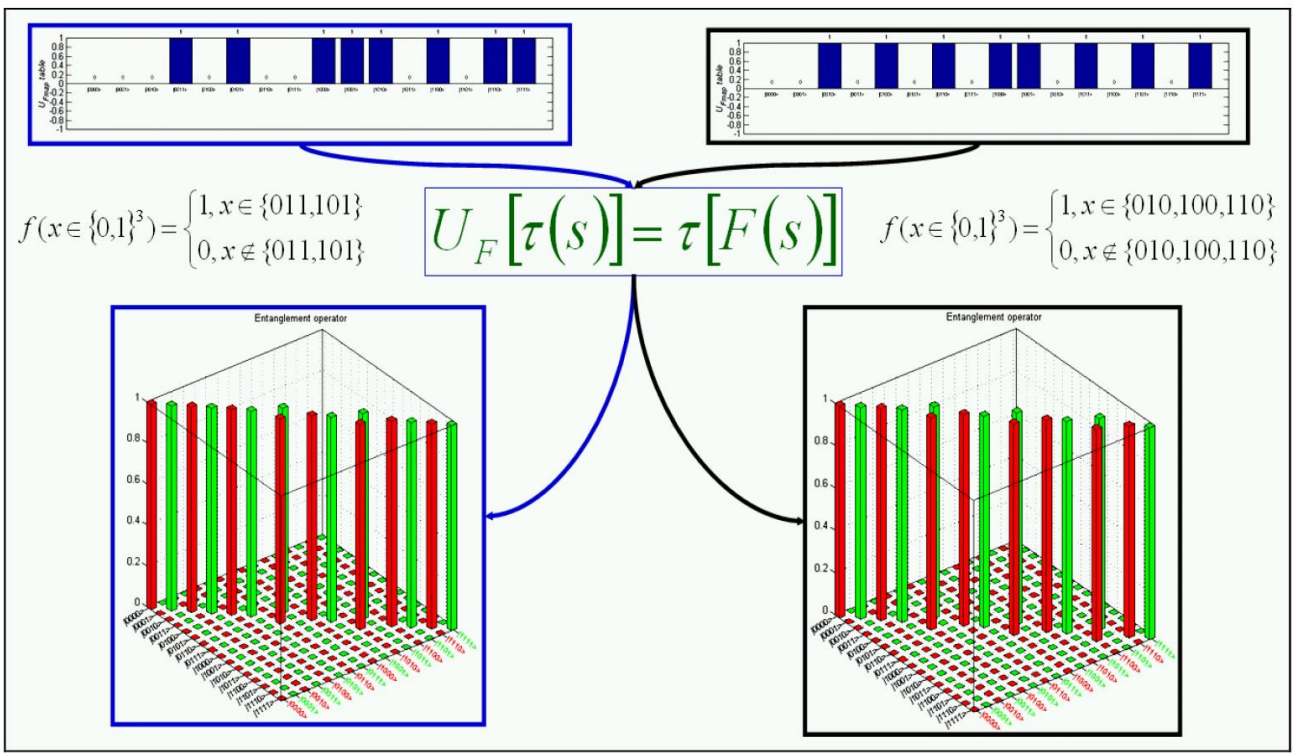

Figure 18: Grover's QA: Step 1.2. Preparation of quantum operators: Entanglement operators for 2 and 3 answers search.

Figure 19 shows the result of interference operator design (step 1.3 of Fig. 12).

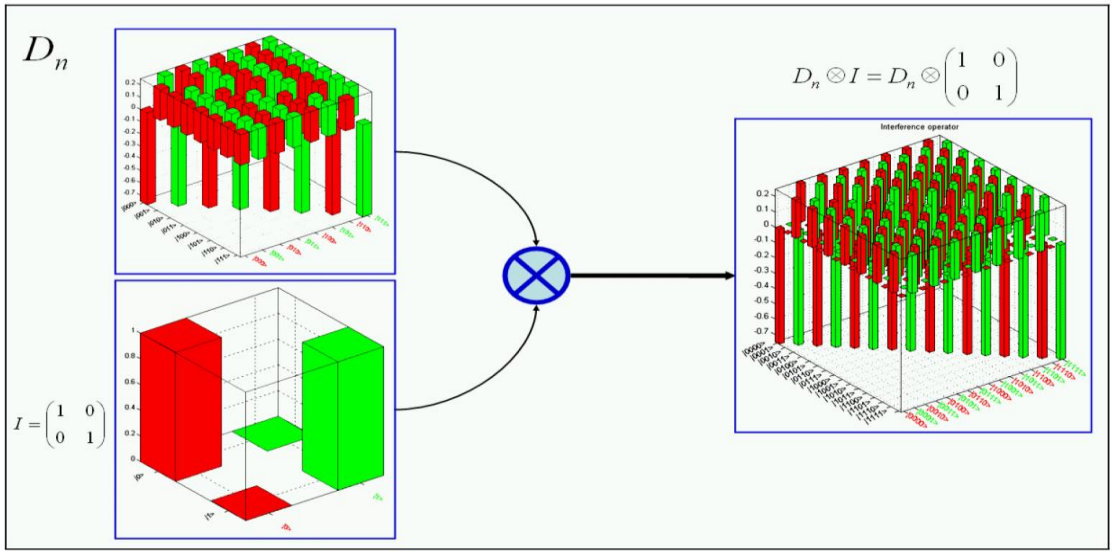
Figure 19: Grover's QA: Step 1.3. Preparation of quantum operators: Interference operator

Comparison between superposition and interference operators in Grover's QAG is shown in Fig. 20

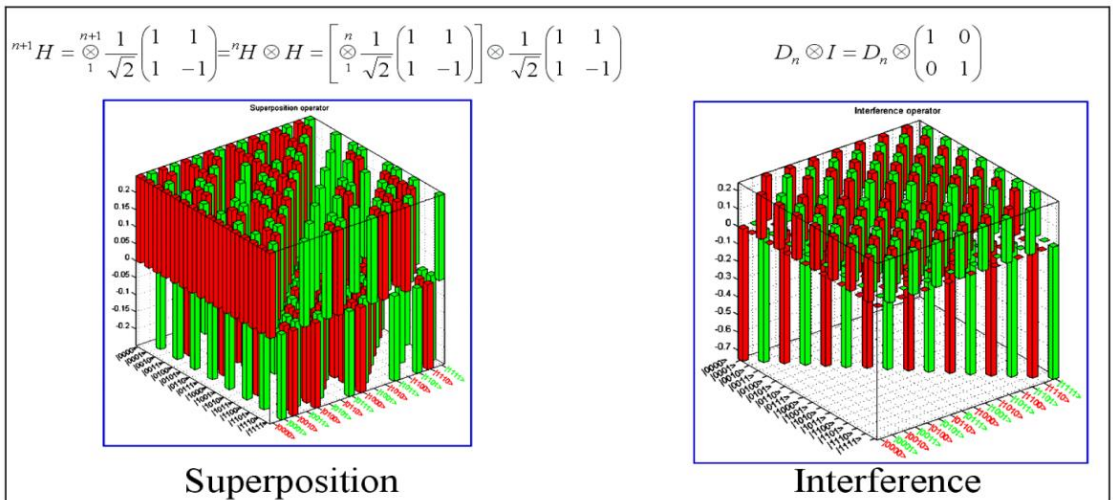
Figure 20: Grover's QA: Superposition and interference operators.

Figure 21 shows the Grover's QAG assembly (step 1.4 of Fig. 2.6).

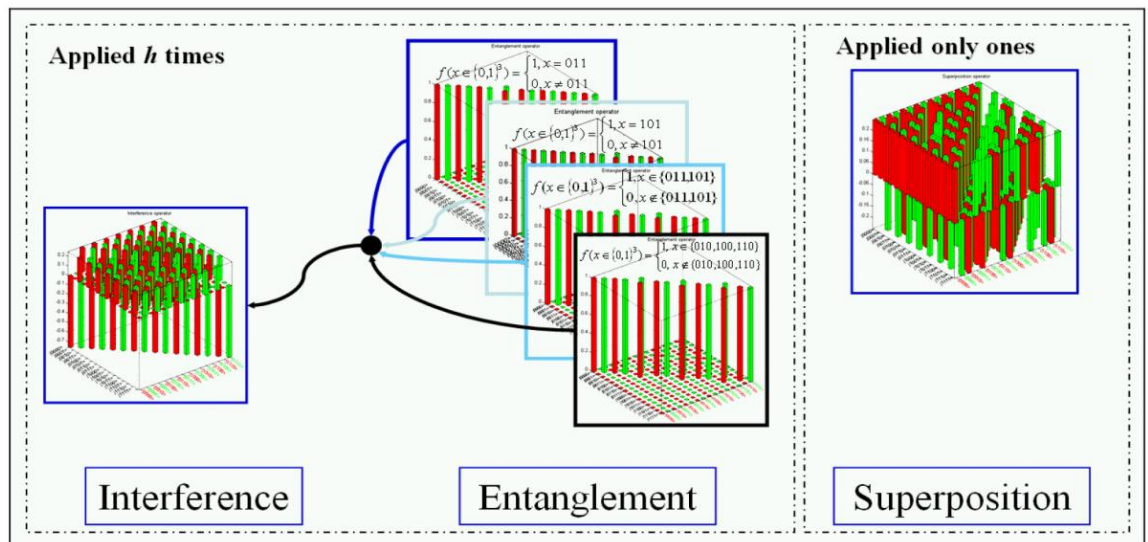
Figure 21: Grover's QA: Step 1.1. Quantum gate assembly.

Figure 22 shows the assembled entanglement and interference operators in gate representation (step 1.4 from Fig. 12).

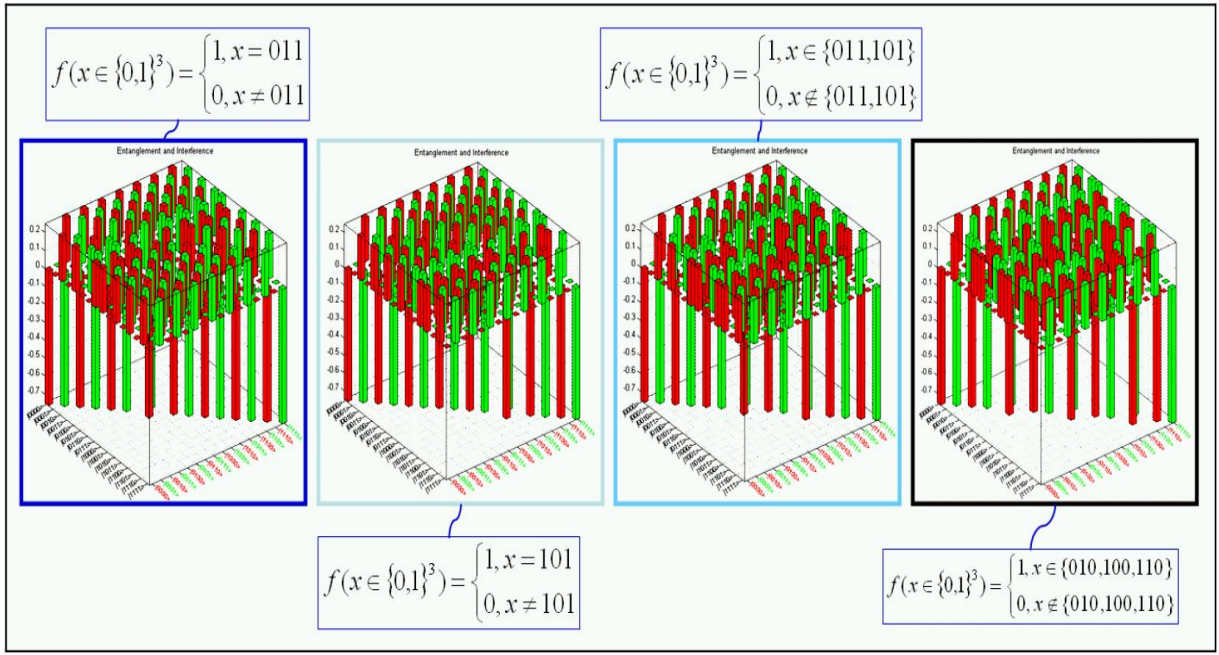

Figure 22: Grover's QA: Step 1.1. Assembled entanglement and interference operators.

Dynamic evolution of successful results of algorithm execution for the first iteration of Grover's QAG for initial qubits state $|0001\rangle$ and different answer search is shown in Fig. 23.

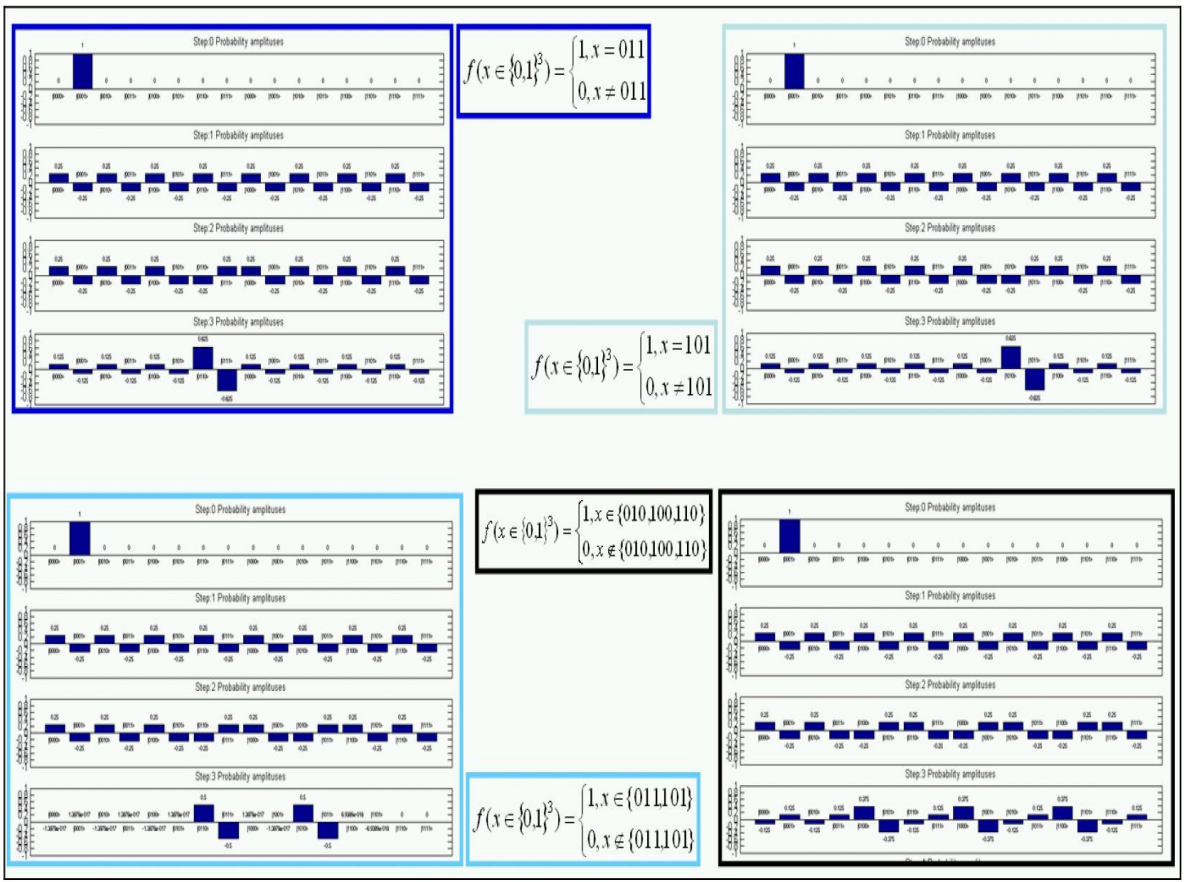

Figure 23: Grover's QA: Algorithm execution. First Iteration.

Figure 24 shows algorithm execution results for Grover's QSA with different number of iterations for successful results with different searching answer number.

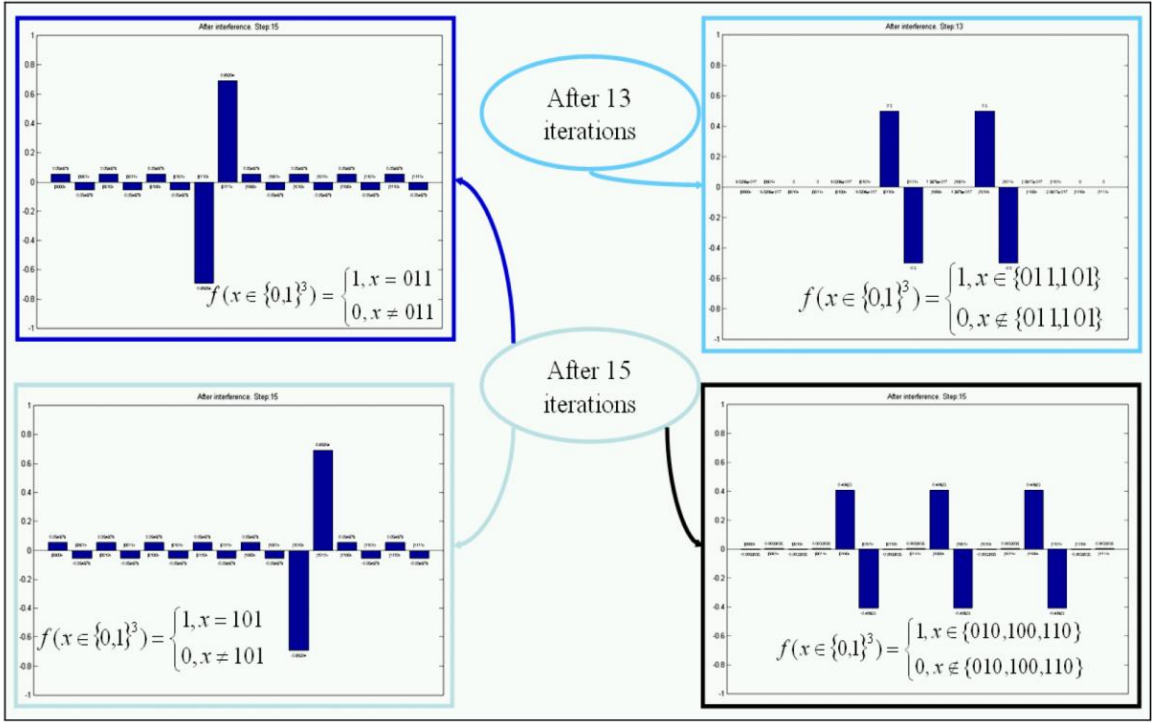

Figure 24: Grover's QA: Step 2. Algorithm execution results.

Algorithm execution 3D dynamics (step 2 of Fig. 12) for the same cases is shown in Fig. 25.

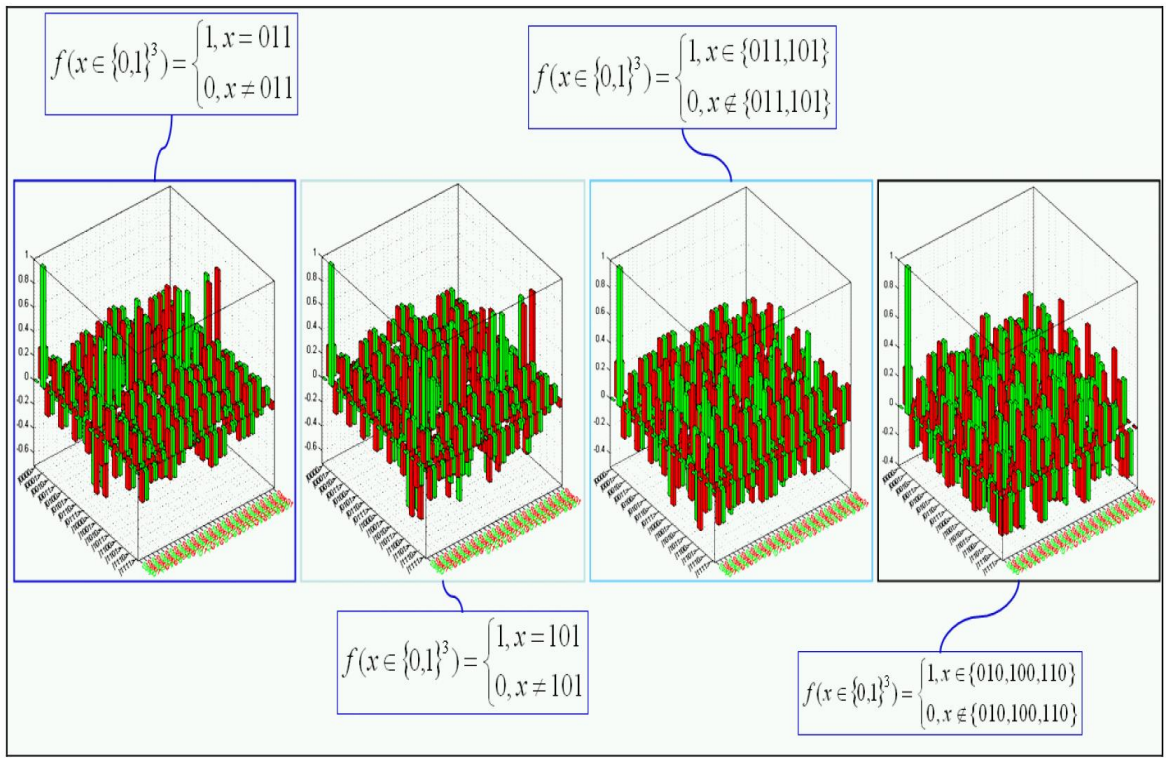

Figure 25: Grover's QA: Step 2 Algorithm execution 3d dynamics: Probability amplitudes.

Figure 26 is a 3D dynamic representation of Grover's QAG probabilities evolution (step 2 of Fig. 12) for different cases of answer search.

Algorithm execution results of Grover's QAG (step 2 of Fig. 12) with different stopping iteration for searching answers are shown in Fig. 25.

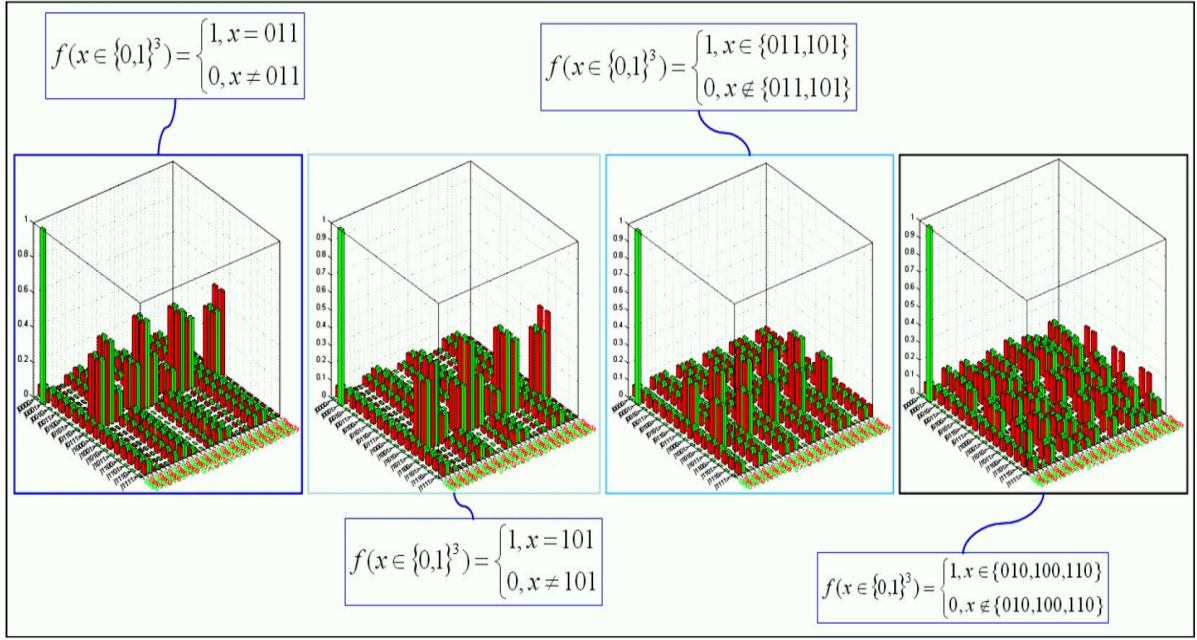
Figure 26: Grover's QA: Step 2 Algorithm execution 3d dynamics: Probabilities.

## 4.3 Interpretation of measurement results in simulation of Grover's QSA-QAG

In the case of Grover's QSA this task is achieved (according to the results of this section) by preparing the ancillary qubit of the oracle of the transformation:

$$U_f : |x, a\rangle \mapsto |x, f(x) \oplus a\rangle$$

in the state $|a_0\rangle = \frac{1}{\sqrt{2}}(|0\rangle - |1\rangle)$. In this case the operator $I_{|x_0\rangle}$ is computationally equivalent to

$$U_F : U_F\left[|x\rangle \otimes \frac{1}{\sqrt{2}}(|0\rangle - |1\rangle)\right] = \left[I_{|x_0\rangle}(|x\rangle)\right] \otimes \frac{1}{\sqrt{2}}(|0\rangle - |1\rangle)$$

$$= \frac{1}{\sqrt{2}}\underbrace{\left[I_{|x_0\rangle}(|x\rangle)\right]}_{\text{Computation result}} \otimes \underbrace{|0\rangle}_{\text{Measurement}} - \frac{1}{\sqrt{2}}\underbrace{\left[I_{|x_0\rangle}(|x\rangle)\right]}_{\text{Computation result}} \otimes \underbrace{|1\rangle}_{\text{Measurement}}$$

and the operator $U_f$ is constructed from a controlled $I_{|x_0\rangle}$ and two one qubit Hadamard transformations.

Figure 27 shows the interpretation of results of the Grover QAG.

| | |
|---|---|
| *If measured basis vector:* | $\underbrace{\left|x_0^1 \cdots x_0^n x_0^{n+1}\right\rangle}_{n+1\,qubits}$ |
| *Consist of:* | $\underbrace{\left|x_0^1 \cdots x_0^n\right\rangle}_{n\,qubits\,of\,computational\,basis} \otimes \underbrace{\left|x_0^{n+1}\right\rangle}_{1\,qubit\,of\,measurement\,basis}$ |
| *Then searched argument was:* | $\left.\underbrace{x_0 = x_0^1 \cdots x_0^n}_{n\,bits}\right\} \Rightarrow \begin{array}{l}\textit{Answer of}\\ \textit{Quantum}\\ \textit{Searching}\end{array}$ |

Figure 27: Grover's QA: Step 2 Result interpretation.

Measured basis vector is computed from the tensor product between the computation qubit results and ancillary measurement qubit. In Grover's searching process, the ancillary qubit does not change during the quantum computing.

As described above, operator $U_f$ is constructed from two Hadamard transformations and the Hadamard transformation $H$ (modeling the constructive interference) applied on the state of the standard computational basis can be seen as implementing a fair coin tossing. Thus, if the matrix

$H = \dfrac{1}{\sqrt{2}}\begin{pmatrix} 1 & 1 \\ 1 & -1 \end{pmatrix}$ is applied to the states of the standard basis, then $H^2|0\rangle = -|1\rangle$, $H^2|1\rangle = |0\rangle$, and therefore $H^2$ acts in measurement process of computational result as a *NOT*-operation, up to the phase sign. In this case, the measurement basis separated with the computational basis (according to tensor product).

The results of simulation are shown in Fig. 28 (a). Figure 28 (b) shows the results of computation on a classical computer.

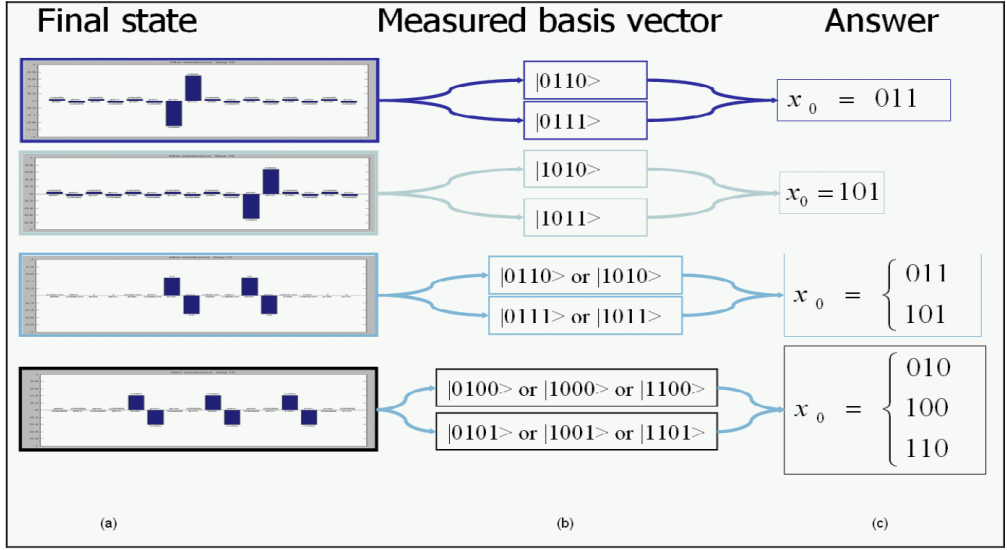

Figure 28: Grover's QA: Results of algorithm.

Figure 27 (b) shows two possibilities:

$$\{|0110\rangle\} = \underbrace{|011\rangle}_{\text{Result}} \otimes \underbrace{|0\rangle}_{\text{Measured qubit}}$$

and

$$\Updownarrow \qquad \Updownarrow$$

$$\{|0111\rangle\} = \underbrace{|011\rangle}_{\text{Result}} \otimes \underbrace{|1\rangle}_{\text{Measured qubit}} \quad .$$

A similar situation is shown in Fig. 27, b.

Figure 27 (b) demonstrate also two searching marked states:

$$\left\{ |0110\rangle = \underbrace{|011\rangle}_{\text{Result}} \otimes \underbrace{|0\rangle}_{\text{Measured qubit}} \quad \text{or} \quad \{|0101\rangle\} = \underbrace{|101\rangle}_{\text{Result}} \otimes \underbrace{|0\rangle}_{\text{Measured qubit}} \right\}$$

and

$$\Updownarrow \qquad\qquad \Updownarrow$$

$$\left\{ |0111\rangle = \underbrace{|011\rangle}_{\text{Result}} \otimes \underbrace{|1\rangle}_{\text{Measured qubit}} \quad \text{or} \quad \{|1011\rangle\} = \underbrace{|101\rangle}_{\text{Result}} \otimes \underbrace{|1\rangle}_{\text{Measured qubit}} \right\}$$

A similar situation is shown for three searching marked states in Fig. 28 (b).

Using a random measurement strategy based on a fair coin tossing in the measurement basis $\{|0\rangle, |1\rangle\}$ one can independently receive with certainty the searched marked states from the measurement basis result. Figure 28 (c) show accurate results of searching of corresponding marked states. Final results of interpretation for

Grover's algorithm are shown in Fig. 26. The measurement results based on a fair coin tossing measurement are shown in Fig. 28 (c).

Figure 28 (c) shows that for both possibilities in implementing a fair coin tossing type of measurement process the search for the answer are successful and demonstrate the possibility of the effectiveness of quantum algorithm simulator realization on classical computer.

Related problems of QA classical simulation in [6-15] discussed.

# Conclusions

- General approach to design of quantum algorithm gates is described.
- Gate-based quantum algorithm representation for effective simulation on computer with classical architecture demonstrated.
- Grover's quantum search algorithm is explained in detail along with implementations on a local computer simulator.